\documentclass[sigconf]{acmart}
\usepackage{graphicx}
\usepackage{subcaption} 
\usepackage{tikz}
\AtBeginDocument{%
  }

\setcopyright{acmlicensed}
\copyrightyear{2018}
\acmYear{2018}
\acmDOI{XXXXXXX.XXXXXXX}
\acmConference[Conference acronym 'XX]{Make sure to enter the correct
  conference title from your rights confirmation email}{June 03--05,
  2018}{Woodstock, NY}
\acmISBN{978-1-4503-XXXX-X/2018/06}

\begin{document}

\title{GenAIReading: Augmenting Human Cognition with Interactive Digital Textbooks Using Large Language Models and Image Generation Models}

\author{Ryugo Morita}
\orcid{}
\affiliation{%
  \institution{Hosei University {\&} DFKI GmbH}
  \city{Tokyo}
  \country{Japan}
}
\email{ryugo.morita.7f@stu.hosei.ac.jp}

\author{Ko Watanabe}
\orcid{0000-0003-0252-1785}
\affiliation{%
  \institution{DFKI GmbH}
  \city{Kaiserslautern}
  \country{Germany}
}
\email{ko.watanabe@dfki.de}

\author{Jinjia Zhou}
\orcid{0000-0002-5078-0522}
\affiliation{%
  \institution{Hosei University}
  \city{Tokyo}
  \country{Japan}
}
\email{jinjia.zhou.35@hosei.ac.jp}

\author{Andreas Dengel}
\orcid{0000-0002-6100-8255}
\affiliation{%
  \institution{DFKI GmbH}
  \city{Kaiserslautern}
  \country{Germany}
}
\email{andreas.dengel@dfki.de}

\author{Shoya Ishimaru}
\orcid{0000-0002-5374-1510}
\affiliation{%
  \institution{Osaka Metropolitan University}
  \city{Osaka}
  \country{Japan}
} 
\email{ishimaru@omu.ac.jp}

\renewcommand{\shortauthors}{Morita and Watanabe et al.}

\begin{abstract}
  Cognitive augmentation is a cornerstone in advancing education, particularly through personalized learning. 
  However, personalizing extensive textual materials, such as narratives and academic textbooks, remains challenging due to their heavy use, which can hinder learner engagement and understanding.
  Building on cognitive theories like Dual Coding Theory—which posits that combining textual and visual information enhances comprehension and memory—this study explores the potential of Generative AI (GenAI) to enrich educational materials. 
  We utilized large language models (LLMs) to generate concise text summaries and image generation models (IGMs) to create visually aligned content from textual inputs. 
  After recruiting 24 participants, we verified that integrating AI-generated supplementary materials significantly improved learning outcomes, increasing post-reading test scores by 7.50\%. 
  These findings underscore GenAI's transformative potential in creating adaptive learning environments that enhance cognitive augmentation.
\end{abstract}



\begin{teaserfigure}
  \centering
  \includegraphics[width=0.9\textwidth]{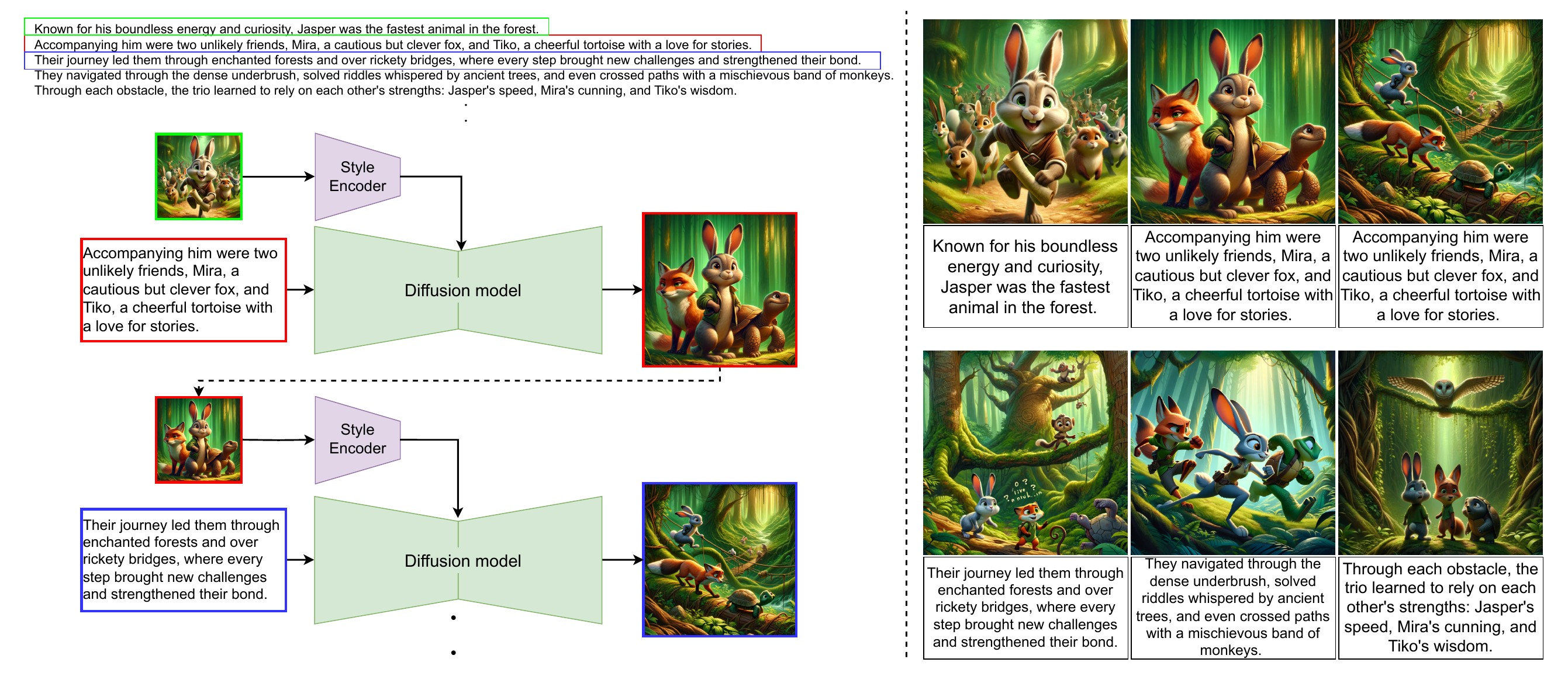}
  \caption{The model takes both a provided sentence and a preceding image as inputs to comprehend the narrative context and stylistic elements.
  It then generates a corresponding story image that aligns with the input sentence and the style inferred from the previous image. The images displayed on the right side demonstrate various outcomes the model produces based on different input sentences.}
  \label{fig:overflow}
\end{teaserfigure}

\received{20 February 2007}
\received[revised]{12 March 2009}
\received[accepted]{5 June 2009}

\maketitle

\section{Introduction}
Cognitive augmentation~\cite{schmidt2017augmenting, kunze2023cognitive, clinch2023augmented} is a pivotal concept in advancing human cognition through technology.
\citet{schmidt2017augmenting} emphasize how interactive technologies—such as remote collaboration tools, mobile computing, and machine learning—have democratized cognitive enhancement, making it widely accessible. 
Similarly, \citet{clinch2023augmented} highlights the transformative role of emerging wearables, advanced sensing modalities, and portable neuroimaging technologies in further enhancing cognitive capabilities.
Together, these innovations demonstrate how technology revolutionizes human cognition, fostering intellectual growth and performance.

In educational psychology, retaining extensive textual materials, such as academic textbooks and narratives, remains a significant challenge~\cite{ishimaru2018hypermind}. 
While these materials are essential for building foundational knowledge, their text-heavy formats can hinder engagement, particularly in the digital age, where visual integration has become vital to improving educational experiences.
Technological advancements—such as cameras, videos, and digital platforms—have expanded the role of visual content in promoting effective learning strategies. 
Paivio's Dual Coding Theory posits that encoding information in textual and visual formats enhances memorization and comprehension~\cite{paivio1991dual}. 
Mayer's research further supports this theory, showing that text integrated with visual aids significantly improves learning outcomes~\cite{mayer2003nine}.

However, the widespread adoption of visual aids in educational materials remains limited due to the significant resources and expertise required to design high-quality, contextually relevant visuals.
Traditional printed textbooks also face physical constraints that restrict the inclusion of supplementary visual content.
These limitations have led educators and publishers to prioritize text-heavy formats, which often fail to address the diverse needs of learners who benefit from multimodal content~\cite{renninger2015interest}.

We propose leveraging Generative AI (GenAI) to bridge this gap and enrich educational materials with adaptive, personalized content for human cognitive augmentation.
Figure~\ref{fig:overflow} represents the overall idea of our application.
Our study investigates the impact of AI-generated supplementary materials—text summaries, images, and image summaries—on learners' comprehension and retention.
We created concise text summaries using Text Generation AI (TGenAI), while Image Generation AI (IGenAI) converted sentences into corresponding images.
We further developed a novel \textit{Summary Image Selector} to intelligently curate the five most relevant images, ensuring alignment with the narrative flow.

Our empirical study involved diverse participants and utilized eye-tracking data to analyze engagement with AI-generated materials. 
The results revealed that incorporating AI-generated text summaries, images, and image summaries improved post-reading test scores by 1.25\%, 4.58\%, and 7.50\%, respectively. 
This demonstrates the potential of AI-driven materials to enhance learning outcomes significantly.
Additionally, we observed correlations between post-reading test scores and learners' preferences for text or image materials. 
Eye-tracking data highlighted that personalized educational content tailored to learners' cognitive profiles can optimize learning experiences, providing further evidence of AI's potential to address diverse learning needs.

This research lays the groundwork for developing AI-driven educational tools that automatically generate adaptive learning materials. 
Our framework can significantly enhance learning outcomes by transforming how content is created and tailored.
Future research should expand on these findings across diverse populations and explore real-world applications of these tools in educational settings.
Our key contributions are as follows:

\begin{itemize} 
  \item[C1] We present a novel framework integrating AI-generated text summaries and images into educational materials, demonstrating how Generative AI can improve comprehension and retention. The innovative \textit{Summary Image Selector} method aligns visual content with the narrative.
  \item[C2] Through a comprehensive empirical study supported by eye-tracking data, we evaluate the effectiveness of AI-generated materials across different learner preferences, highlighting their potential to enhance educational outcomes.
  \item[C3] We provide design guidelines for AI-driven educational tools, emphasizing optimizing visual content to accommodate diverse cognitive styles and learning preferences.
\end{itemize}

\section{Related Work and Background}
In this section, we introduce prior work on use of generative AI in human-computer interaction, and human memory and comprehension augmentation.

\begin{figure*}[t!]
  \centering
  \includegraphics[width=0.85\textwidth]{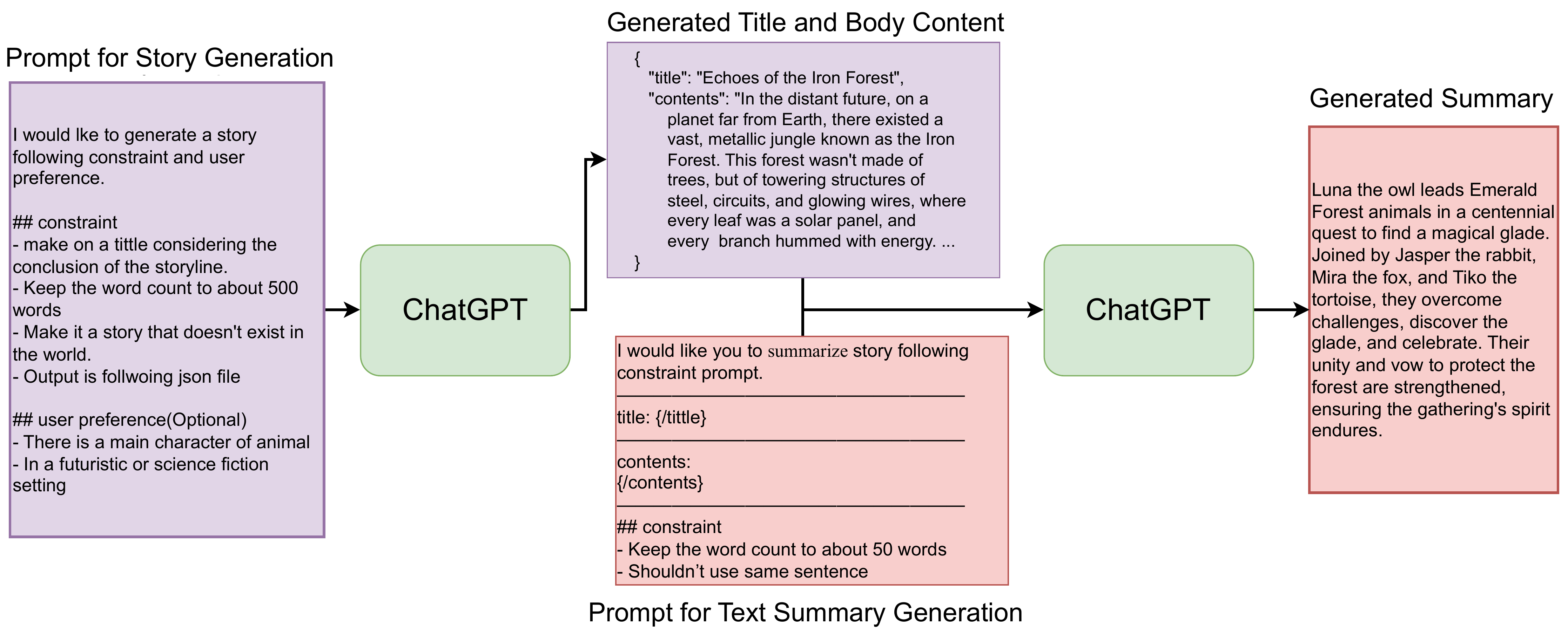}
  \caption{Architecture of the generation flow of the story text summary using LLMs (ChatGPT). The input consists of the generated story from the story generation phase and constraint prompts, which guide the summary generation. The constraint prompts control parameters such as word count, ensuring that the summary is concise and adheres to the specified length and content requirements for effective summarization.}
  \label{fig:text_gen}
\end{figure*}

\subsection{Generative AI in Human-Computer Interaction}
\label{ssec:genai_hci}
Generative AI, encompassing text and image generation technologies, has significantly impacted various domains within Human-Computer Interaction (HCI). Since the advent of Transformers~\cite{vaswani2017attention}, Transformer-based Large Language Models (LLMs) have become prevalent tools in HCI applications.

LLMs have been utilized in diverse fields, including personalized marketing and targeted advertising~\cite{feizi2023online, yang2024somonitor}, creative arts such as music generation~\cite{yuan2024chatmusician}, and the aesthetic evaluation of poetry and storytelling~\cite{vear2024jess}. 
In visual arts, LLMs assist in creative decision-making, like analyzing color theory in interior design~\cite{hou2024c2ideas}. 
In education, LLMs support the ``Learning by Teaching'' paradigm, where students teach AI agents to reinforce their understanding while receiving feedback~\cite{ali2023supporting, markel2023gpteach}. 
They also aid in study planning and organizational support~\cite{sun2024adaplanner}, vocabulary learning~\cite{yamaoka2022experience}, and provide writing assistance~\cite{freire2023may, park2024promise}.
While LLMs have seen extensive adoption in HCI, the rise of diffusion-based image generation models introduces new possibilities and challenges. 
These models have enabled significant advancements in visual content creation, particularly in user interface design and interactive experiences~\cite{liu2022design, wang2024promptcharm}.

Parallel to LLMs, diffusion-based image generation models have emerged, introducing new possibilities for visual content creation in HCI~\cite{liu2022design, wang2024promptcharm}. 
These models have facilitated advancements in user interface design and interactive experiences. However, their adoption in educational contexts remains limited, primarily due to the labor-intensive nature of prompt engineering~\cite{liu2022design}.
Recent research has begun addressing these challenges by integrating text and image generative AI (T+IGenAI). 
Leveraging LLMs to generate prompts for image generation reduces user burden and enhances creative workflows~\cite{petridis2024promptinfuser}. 
Studies have explored using LLMs to assist in video editing~\cite{wang2024lave} and facilitating collaborative creation between AI and humans~\cite{fan2024contextcam}. 
Additionally, Generative AI has been employed to create fairy tales that combine textual and visual elements to evoke emotional responses~\cite{harde2024generative}.

Despite these advancements, there is a noticeable gap in research demonstrating the practical effectiveness of Generative AI in real-world educational applications. Our research aims to fill this gap by leveraging text and image generation AI to create visually enriched educational content. By generating content aligned with individual students' interests and preferences, we seek to increase engagement and motivation in the learning process.

\begin{figure*}[t!]
  \centering
  \includegraphics[width=0.8\textwidth]{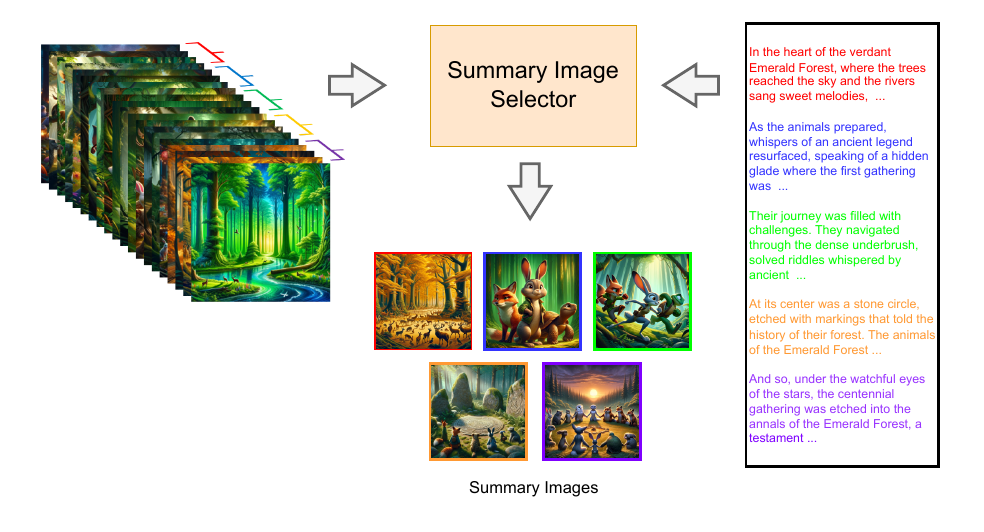}
  \caption{Architecture of the selection flow of the summary image selector. The input includes the story and the generated images, which are processed to select five key summary images. The text and images are segmented and fed into the Summary Image Selector to calculate the highest similarity score in each segment, which is chosen as the summary image.}
  \label{fig:img_summary}
\end{figure*}

\subsection{Human Memory and Comprehension Augmentation}
Understanding human memory and comprehension is crucial for enhancing educational outcomes. Research indicates that both textual and visual information play critical roles in memory retention, and combining these modalities can enhance learning.

Textual information involves complex cognitive processes. 
The presentation of information significantly influences reading comprehension~\cite{pramod2021text} and memory retention~\cite{elagroudy2022memory}. 
For instance, coherent and well-structured text aids in better understanding and recall~\cite{kintsch2005comprehension}. 
Elaborative encoding, which entails processing the meaning of information and linking it to existing knowledge, improves memory retention~\cite{craik1972levels}.
Visual information is processed differently and often has a substantial impact on memory and comprehension.
According to Paivio's Dual Coding Theory, information encoded both verbally and non-verbally enhances comprehension~\cite{paivio1991dual}.
This is supported by the picture superiority effect, where images are remembered better than words~\cite{nelson1976picture}. 
\citeauthor{shepard1967recognition} demonstrated that people could recall images accurately even after long delays, indicating robust visual memory~\cite{shepard1967recognition}.
Integrating text and visual information can significantly enhance memory retention. 
\citeauthor{mayer2003nine} found that combining words and pictures leads to better learning outcomes than using either alone~\cite{mayer2003nine}. 
This multimedia learning effect is particularly strong when visuals directly relate to the text, providing contextual support that aids understanding and retention. 
\citeauthor{carney2002pictorial} showed that pictorial illustrations can improve memory for textual information by providing visual anchors that assist in retrieval~\cite{carney2002pictorial}.
Personal interest and the relevance of information also influence memory retention. Individuals are more likely to remember personally interesting or relevant information. 
\citeauthor{renninger2015interest} found that interest enhances cognitive processing, leading to better memory retention~\cite{renninger2015interest}. 
Similarly, \citeauthor{schiefele1991interest} demonstrated that students interested in a topic recalled more information and had better comprehension than less interested peers~\cite{schiefele1991interest}.

Our research leverages these insights by integrating AI-generated supplemental content to enhance educational materials. 
By combining textual summaries and relevant images generated by AI, we aim to improve memory retention and comprehension, tailored to individual learner preferences.

\section{Methodology}
This study is structured around two key phases: the Text Generation Phase (TGP) and the Image Generation Phase (IGP). 
The TGP comprises two main components: story generation and text-summary generation. 
The IGP, on the other hand, includes the sentence-image generation and the summary-image selector. 
In the following sections, we explain the processes and methodologies involved in each phase.

\subsection{Approach for the Text Generation (TGenAI)}
\label{sec:text_generation}
Inspired by the StoryPrompt~\cite{fan2024storyprompt}, we utilized the ChatGPT to generate story datasets. 
The ChatGPT is an advanced large language model based on GPT-3~\cite{winata2021language}, designed to generate coherent and contextually relevant text across a wide range of domains, using a combination of supervised learning and reinforcement learning from human feedback (RLHF)~\cite{ouyang2022training}.
Firstly, we create the story within the Story Generation section. 
Once the story is generated, it is distilled into a concise summary in the Summary Generation section. 
This summary not only provides a quick reference for the story but also plays a crucial role in guiding the subsequent Image Generation phase, ensuring that the visual content accurately reflects the core elements of the story.

\subsubsection{Story Generation} 
\label{sec:story_generation}
Figure~\ref{fig:text_gen} shows the input for this section consists of instruction, constraint, and preference prompts, and the output is a generated story that adheres to these parameters.
The constraint prompt imposes specific limitations, such as word count restrictions and the requirement that the generated content must be completely original, ensuring no prior existence in any form.
The preference prompt allows story customization based on user-defined elements, including preferred animals (e.g., rabbits, foxes) and story genres (e.g., adventure, science fiction). 
Users can also provide the title of a favorite story to guide the generation of a personalized story. 
Following the generation of the story, morphological and dependency parsing were conducted to analyze the grammatical structure and relationships within the text. These analyses ensure that the visual elements generated in the subsequent phase accurately represent the story, maintaining consistency in character and thematic elements.

\subsubsection{Summary Generation.} 
The input for this section consists of the generated story from Section.\ref{sec:story_generation} and constraint prompt, and the output is a concise summary.
The constraint prompts control of various aspects of the summary generation, including word count limitations, ensuring that the summary is concise and aligned with the specified length and content requirements.

\begin{figure*}[t!]
  \centering
  \includegraphics[width=0.85\textwidth]{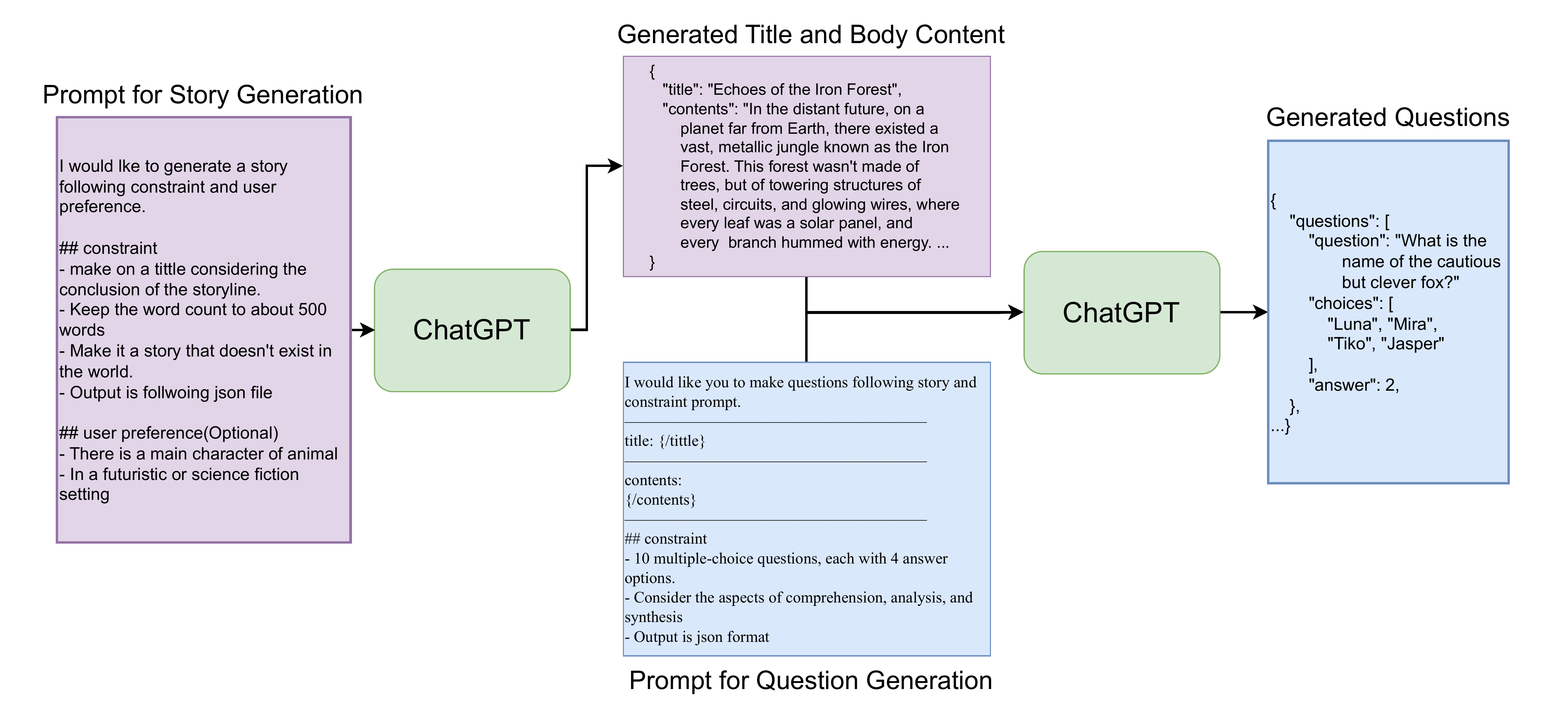}
  \caption{Architecture of the generation flow of the questions using LLMs (ChatGPT). The input is generated story from the story generation phase to tailor questions to align with the story content and constraint prompts. The prompts define question types, such as multiple-choice or open-ended, and determine the focus areas like numerical values or narrative comprehension, ensuring the output is formatted appropriately for further use.}
  \label{fig:question_generation}
\end{figure*}

\subsection{Approach for the Image Generation (IGenAI)}
\label{sec:img_generation}
Firstly, we create the story image from each sentence from the generated story.
Then, the generated images input the Summary Image Selector to obtain the summary images. 

\subsubsection{Story Image Generation.}
\label{sec:story_img_generation}
In this section, we use the diffusion-based text-to-image model DALL-E~\cite{ramesh2021zero} to generate highly detailed and contextually relevant images based on the given prompts.
As shown in Figure~\ref{fig:overflow}, the inputs typically include the generated story and the preceding image to ensure stylistic continuity. 
In the initial image generation process, where no prior image exists, only the textual input is used as the basis for generating the first image in the sequence. The aim is to generate a series of images that match the sentences and maintain a consistent style throughout.

First, the generated story was segmented into individual sentences. 
Each sentence, the results of the dependency parsing, and the summary were used as input to generate corresponding images.  
This process involves sentence-to-image generation, where the dependency parsing results - including character and style information - are continuously integrated into the prompts. 
This ensures that the generated images consistently reflect the narrative coherence and character continuity.
Furthermore, a reference image from the previous output was incorporated as input to maintain visual consistency across the sequence of images. By including the story summary, the model could generate images that grasped the overall storyline and conclusion, enhancing the narrative's visual representation. 
We carry out this process regressively to obtain a series of images.

\subsubsection{Summary Image Selection.}
\label{sec:summary_img}
In this section, we use the CLIP to select the summary images. 
The CLIP (Contrastive Language-Image Pre-training)~\cite{radford2021learning} is a model that aligns images and text in a shared embedding space. 
By jointly training on a large dataset of images paired with textual descriptions, CLIP learns to associate textual and visual information effectively. 
It allows it to evaluate the similarity between a text and an image, making it particularly useful for tasks that require matching or retrieving images based on textual input.

As illustrated in Figure~\ref{fig:img_summary}, this section requires the story and the generated story images as inputs to select five summary images as outputs.
The story was divided into five segments based on the maximum token limit of the CLIP text encoder, ensuring that each segment is processed within the model's optimal capacity for accurate similarity calculation.
The corresponding set of images generated in Section~\ref{sec:story_img_generation} was also segmented according to their respective sentences. 
Each segmented text and image was input into the CLIP encoder to obtain sentence and image vectors. 
The cosine similarity between these vector spaces was calculated, and the image with the highest similarity score for each segment was selected as the summary image for that part.
The summary image was determined as follows:
\[
\text{CLIP-S}(c_i, v_j) = w \cdot \max(\cos(c_i, v_j), 0)
\]
where \( w = 2.5 \), \( \cos(c_i, v_j) \) denotes the cosine similarity between the textual embedding \( c_i \) and the visual embedding \( v_j \). The weight \( w = 2.5 \) was chosen based on the findings of CLIP-Score~\cite{hessel2021clipscore}, demonstrating that this value optimizes the balance between text-image alignment and visual representation. 
This weight emphasizes the highest similarity scores, ensuring that the most representative images are selected as summary images.
To select the summary image for each segment \( i \), we compute:
\[
\text{summary\_image}_i = \arg\max_j \left( \text{CLIP-S}(c_i, v_j) \right) \quad (i = 1, \dots, 5, \, j = 1, \dots, \frac{n}{5})
\]
Based on the highest vector similarity, we select the most representative image as a summary image for each segment.

% \begin{table}[t!]
%   % \renewcommand{\arraystretch}{1.5}
%   \centering
%   \caption{List of four reading conditions prepared in this study.}
%   \begin{tabular}{l|l|l}
%     \hline
%     ID & Condition       & Detail  \\
%     \hline
%     C1 & Baseline        & Participant read a document without any Generative AI augmentation. \\
%     C2 & IGenAI Image    & Participant read a document with Generative AI image augmentation. \\ 
%     C3 & TGenAI Summary  & Participant read a document with Generative AI text summary augmentation. \\
%     C4 & IGenAI Summary  & Participant read a document with Generative AI image summary augmentation. \\
%     \hline
%   \end{tabular}
%   \label{table:reading_conditions}
% \end{table}
\begin{figure*}[t!]
  \centering
  \begin{subfigure}[b]{0.455\textwidth}
      \centering
      \includegraphics[width=\textwidth]{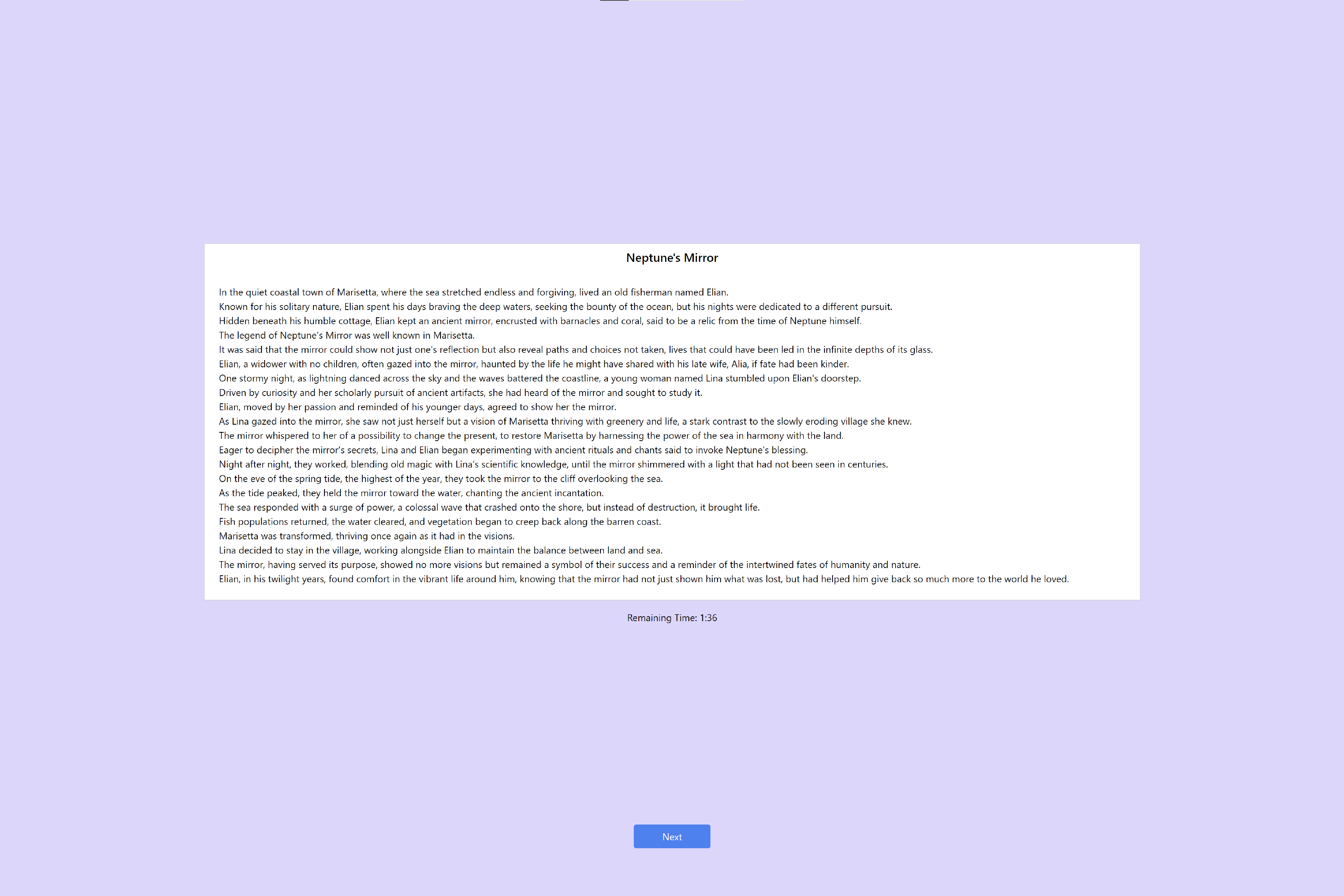}
      \caption{Condition 1: Baseline}
  \end{subfigure}
  \hspace{2mm}
  \begin{subfigure}[b]{0.455\textwidth}
      \centering
      \includegraphics[width=\textwidth]{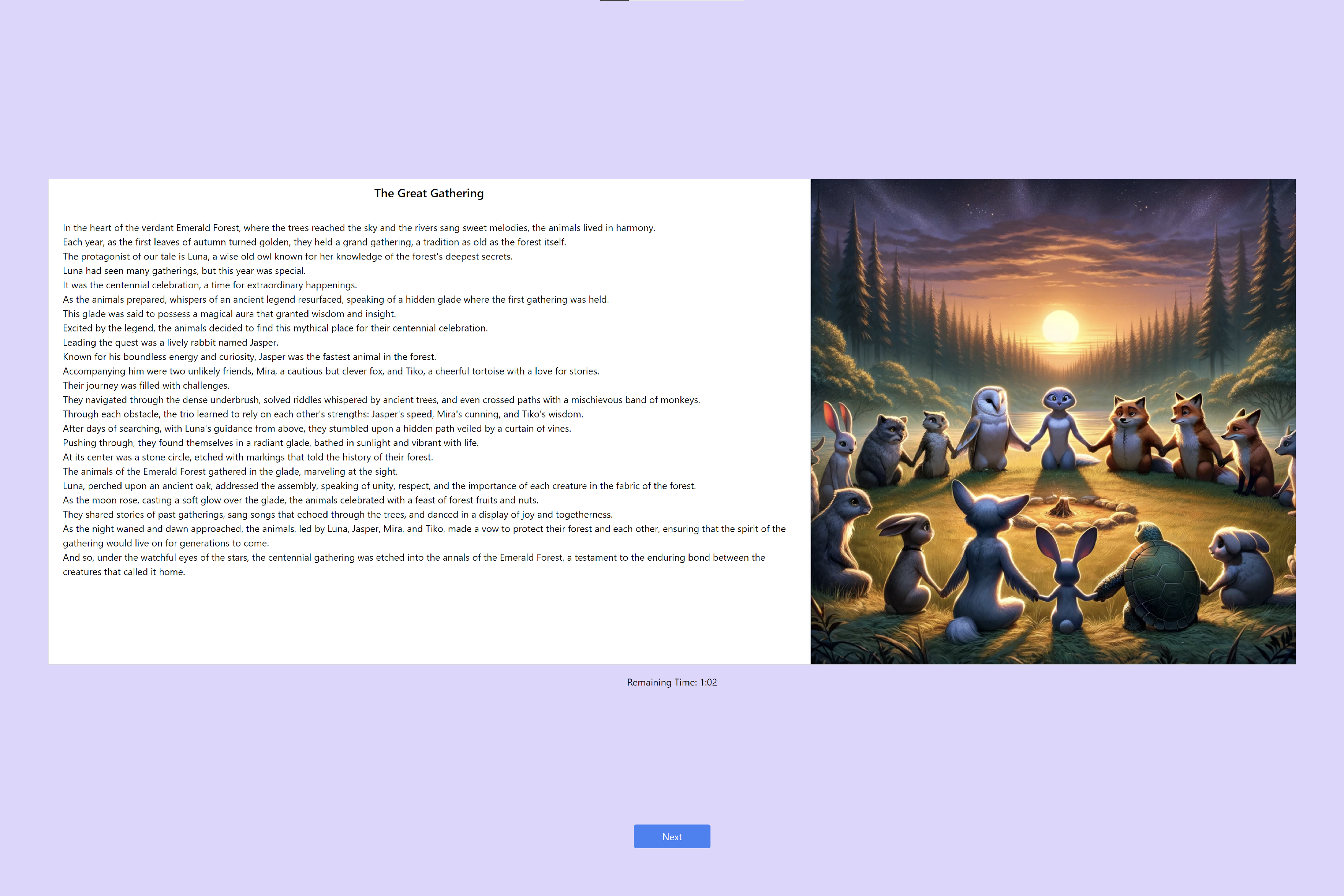}
      \caption{Condition 2: IGenAI Image}
  \end{subfigure}
  \begin{subfigure}[b]{0.455\textwidth}
      \centering
      \includegraphics[width=\textwidth]{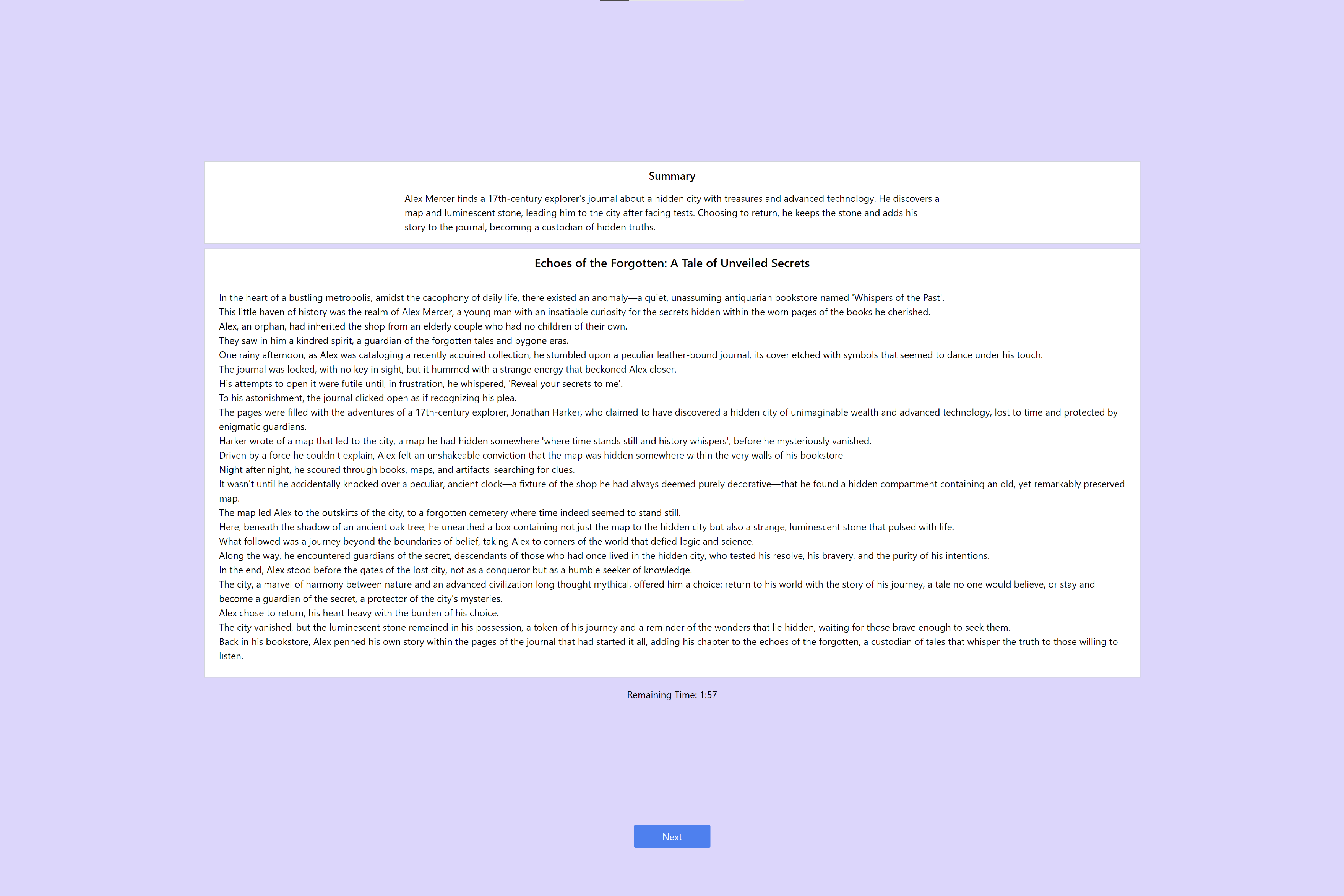}
      \caption{Condition 3: TGenAI Summary}
  \end{subfigure}
  \hspace{2mm}
  \begin{subfigure}[b]{0.455\textwidth}
      \centering
      \includegraphics[width=\textwidth]{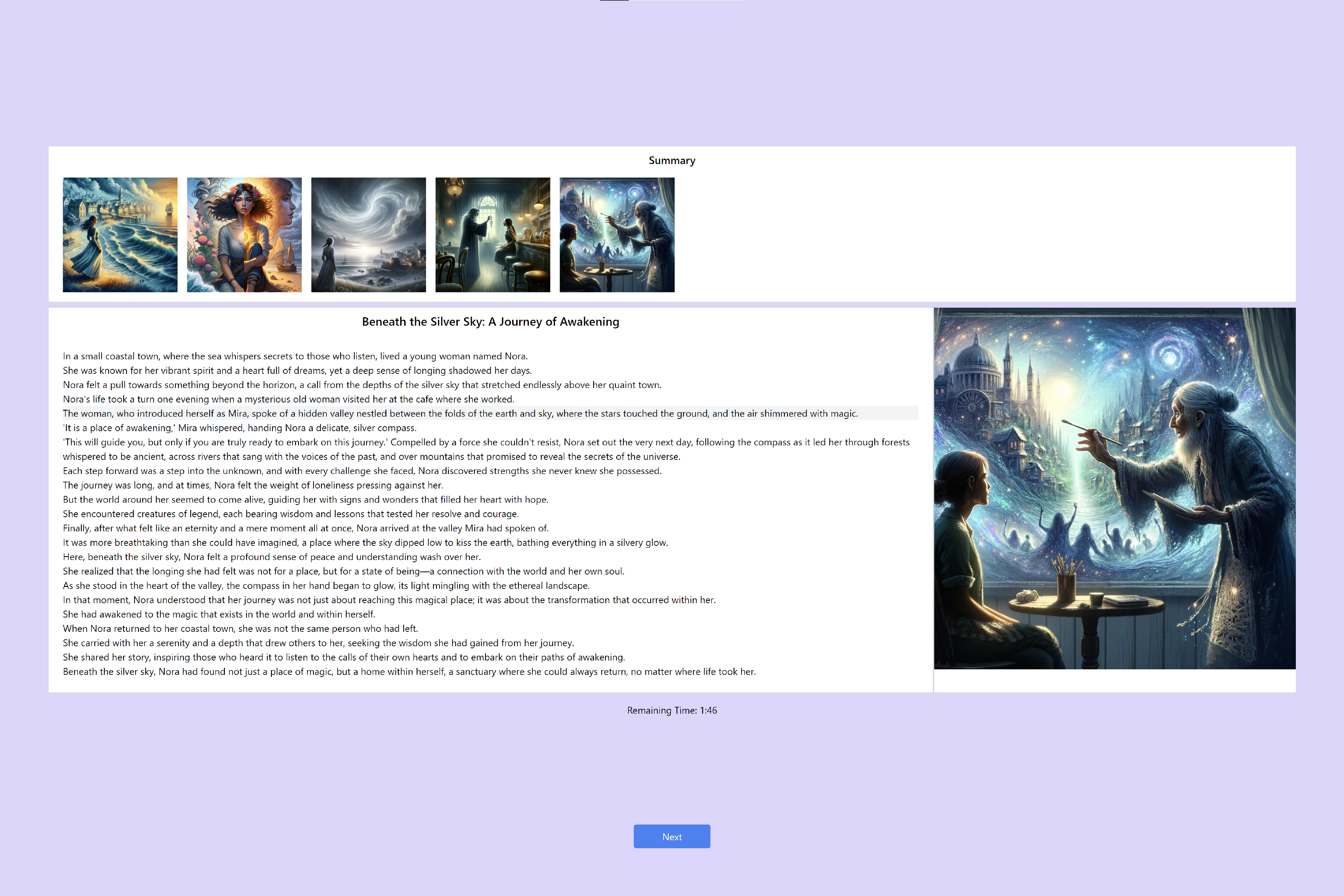}
      \caption{Condition 4: IGenAI Summary}
  \end{subfigure}
  \caption{User interface of the web application showing the four reading conditions: (a) ``Baseline'', (b) ``IGenAI Image'', (c) ``TGenAI summary'', and (d) ``IGenAI Summary''.}
  \label{fig:app}
\end{figure*}

\begin{table*}[t!]
  \centering
  \resizebox{0.9\textwidth}{!}{ % ここでテーブルを幅に合わせて縮小
  \begin{tabular}{l|l|l}
    \hline
    ID & Condition       & Detail  \\
    \hline
    C1 & Baseline        & Participant read a document without any Generative AI augmentation. \\
    C2 & IGenAI Image    & Participant read a document with Generative AI image augmentation. \\ 
    C3 & TGenAI Summary  & Participant read a document with Generative AI text summary augmentation. \\
    C4 & IGenAI Summary  & Participant read a document with Generative AI image summary augmentation. \\
    \hline
  \end{tabular}
  }
  \caption{List of four reading conditions prepared in this study.}
  \label{table:reading_conditions}
\end{table*}

\subsection{Preparation of Questions for Reading Comprehension Evaluation}
\label{sec:question_generation}
Figure~\ref{fig:question_generation} shows the overall generating questions pipeline.
We focus on generating questions that align with the story and constraints prompt, ensuring that the questions are tailored to the content and parameters of the generated story.
The constraint prompts define the types of questions to be generated, including multiple-choice, open-ended, or fill-in-the-blank formats.
Additionally, the prompts determine the specific focus of the questions, such as assessing numerical values, proper nouns, or broader narrative comprehension.
Furthermore, the constraint prompts allow users to specify the desired output format and structure. This ensures the generated questions are organized in a suitable file type or format for subsequent use.

\begin{figure*}[t!]
    \centering
    \includegraphics[width=0.8\textwidth]{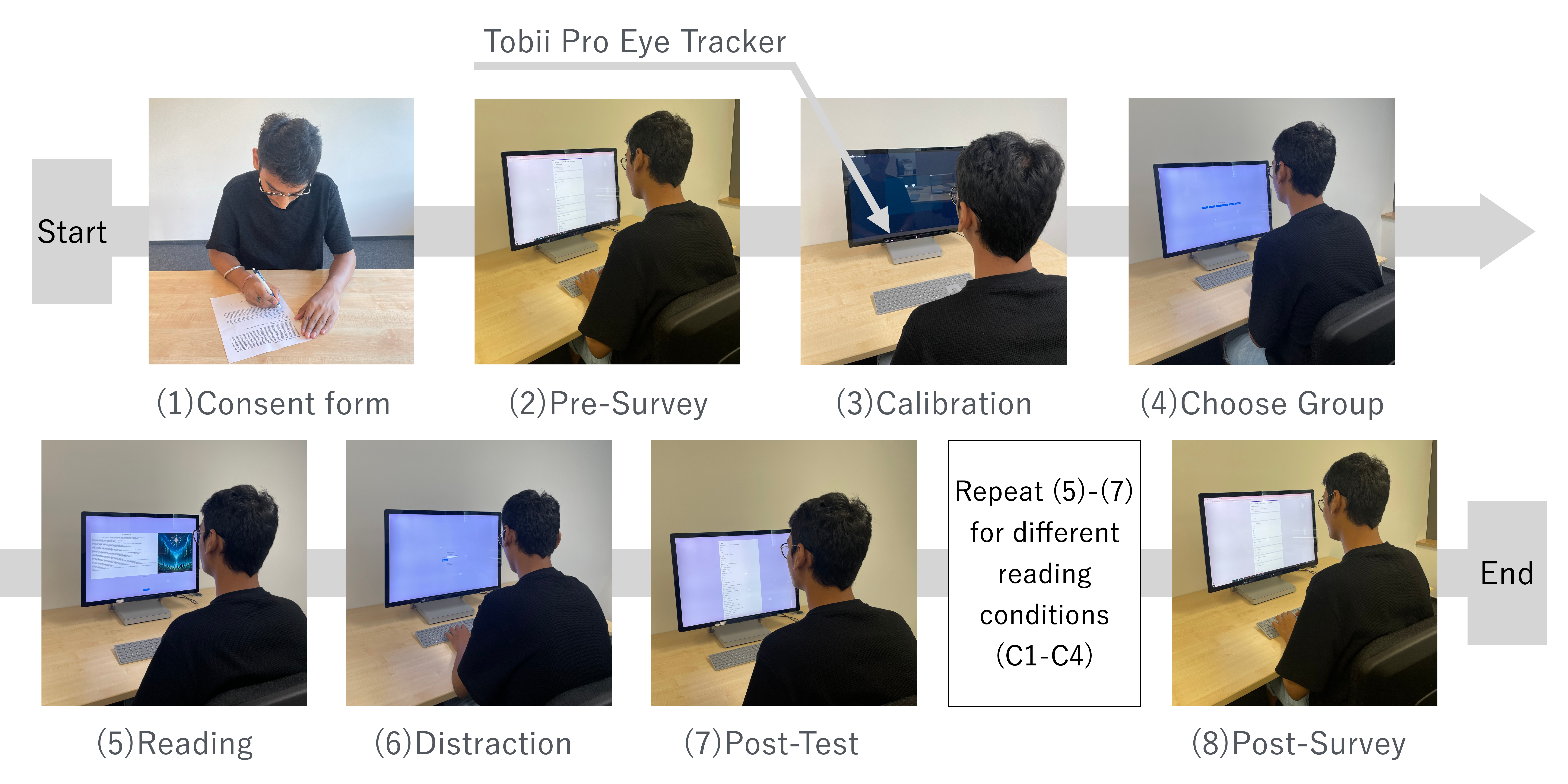}
    \caption{Experiment workflow. Calibration refers to an eye-tracker, the process of estimating the geometric characteristics of a subject's eyes. The post-reading test provides ten questions for evaluating reading comprehension and memory retention of the provided reading conditions.}
    \label{fig:experiment_workflow}
\end{figure*}

\subsection{Web Application}
In this section, we explain how our web application was implemented for the user study.
First, we explain the system workflow of this application.
Then, we explain the log metrics collected using the application.

\subsubsection{Order of Reading Conditions}
\label{subsubsec:group_selection}
One story was randomly selected and designated as the fixed story to assess participants' English language proficiency.
The remaining three stories were divided into six distinct combinations (3! = 6), each assigned to one of the following conditions C2-C4 presented in Table~\ref{table:reading_conditions}.
To present the content consistently and visually organized, an application was developed using React and Node.js, as depicted in Figure~\ref{fig:app}. 
Upon initiating the task, the content for the base condition is centrally displayed on the screen. 
In the C2 condition, the text is presented on the left side of the screen, with the corresponding image displayed on the right. 
In the C3 condition, the time limit was set based on the length of the story, not the summary, to maintain a consistent information load.
In the C4 condition, the image is positioned above the text.

\subsubsection{User Interface of IGenAI}
In the IGenAI page, before the reading task, participants were informed that hovering their mouse over a particular section of the text would trigger the generation of an image related to that specific content. As participants begin reading, whenever they hover over a sentence or phrase, the application immediately generates and displays the corresponding image on the right side of the screen.
If the participant moves the cursor away from the text, the last hovered image remains on the screen, ensuring continuous visual support throughout the reading process. This interactive feature allows participants to receive visual cues in real-time, enhancing their comprehension by providing immediate visual context for their reading text.

\subsubsection{Reading Time Limit}
\label{subsubsec:readin_time_limit}
The application is programmed to automatically transition to the next page once the allotted time limit is reached. 
Reading times were intentionally set to a faster rate of 250 words per minute~\cite{trauzettel2012standardized}, to challenge participants to process and retain information under a more demanding time constraint.

\subsubsection{Log Metrics} 
\label{subsubsection:log_metrics}
In this study, we systematically collected log data and gaze behavior during the reading tasks.

\begin{itemize} 
    \item Duration: The time taken (in seconds) to read the text and complete the accompanying questions.
    \item Group Number: The number of groups the participants are. 
    \item Story Index: The index of the story the participants read.
    \item Distraction Score: The total number of correct responses provided by participants in the distraction task.
    \item Number of Correct Answers: The total number of correct responses provided by participants. 
\end{itemize}

\subsection{Eye-Tracking Data Processing}

\subsubsection{Fixation}
We extracted fixations by grouping nearby eye-tracking gaze points using the Identification by Dispersion-Threshold algorithm, as described by \citet{buscher2008eye}. 
We identify a new fixation when nine consecutive eye-tracking gaze points are detected within 50 pixels of each other. With our eye-tracker sampling at 90 Hz, the nine-point threshold sets the minimum fixation duration to 100 ms. The 50-pixel threshold ensures that the gaze points fall within the same region of interest. To account for measurement noise and small eye movements such as microsaccades, the dispersion threshold is increased to 80 pixels for subsequent points. Thus, gaze points are added to the fixation one at a time as long as they fall within the area determined by the threshold. If a gaze point does not meet the requirement, the algorithm resets and begins counting a new set of nine gaze points to identify a new fixation.

\subsubsection{Areas of Interest (AOI) and AOI Ratio}
The calculated fixation duration and coordinates are used to extract the areas of interest (AOI)~\cite{king2019improving}.
As shown in Figure~\ref{fig:app}, the reading content, generated image, and summary areas are held constant in our user interface.
Our approach allows us to estimate whether the fixation is on the reading content, a generated image, or a summary.
Using the coordinate of the fixation, we apply a threshold to separate the AOI in each content.
Instead of using a fixation duration to evaluate the reading behavior, we use an AOI ratio.
As explained in Section~\ref{subsubsec:readin_time_limit} the reading time duration differs between the document's word count.
Therefore, our experiment focuses on calculating the AOI ratio between the reading content area and the generated image or summary area.

\begin{figure*}[t!]
    \centering
    \begin{subfigure}[b]{0.47\textwidth}
        \centering
        \includegraphics[width=\textwidth]{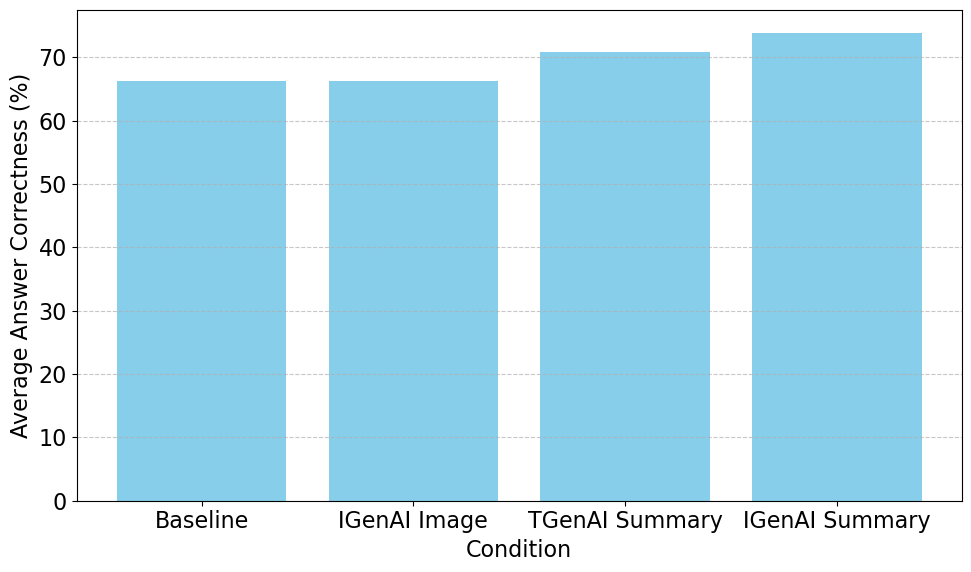}
        \caption{The participants' average score of the post-reading test score}
        \label{fig:zentai1}
    \end{subfigure}
    \begin{subfigure}[b]{0.47\textwidth}
        \centering
        \includegraphics[width=\textwidth]{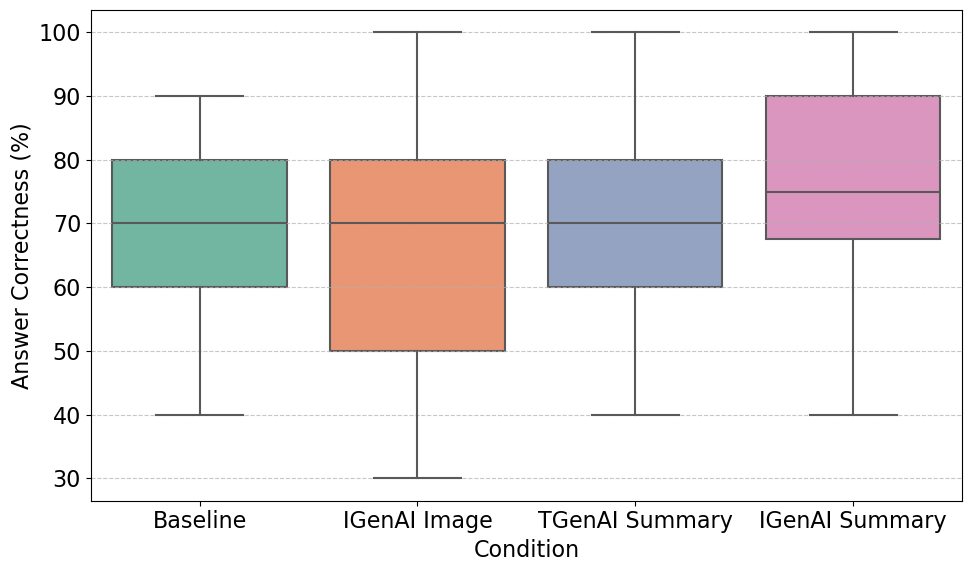}
        \caption{The box-plot of the participants' post-reading test score}
        \label{fig:zentai2}
    \end{subfigure}
    \caption{The comparison of the reading comprehension and memory retention using participants' post-reading test score results in all reading conditions (C1-C4).}
    \label{fig:post_test_score_bargraph}
\end{figure*}

\begin{table*}[t!]
  \renewcommand{\arraystretch}{1.0}
  \centering
  \begin{minipage}{0.45\textwidth}
    \centering
    \caption{List of questions in the pre-survey.}
    \begin{tabular}{l|l}
      \hline
         & Pre-Survey Questions \\
      \hline
      Q1 & Which group are you?  \\
      Q2 & What is your name?    \\
      Q3 & What is your age?     \\
      Q4 & Where are you from? (Nationality) \\
      Q5 & What is your gender? \\
      Q6 & What is your occupation? \\
      Q7 & How long have you been using English? \\
      Q8 & What do you think about your English skills? \\
      Q9 & Are you familiar with LLMs? \\
      Q10 & Are you familiar with IGMs? \\
      \hline
    \end{tabular}
    \label{table:pre_survey}
  \end{minipage}%
  \hfill
  \begin{minipage}{0.45\textwidth}
    \centering
    \caption{List of questions in the post-survey.}
    \begin{tabular}{l|l}
      \hline
         & Post-Survey Questions            \\
      \hline
      Q1 & To what extent are you familiar \\
         & with Large Language Models (LLMs)? \\
      Q2 & To what extent are you familiar \\
         & with Image Generation Models (IGMs)? \\
      Q3 & Which textbook (reading condition) \\
         & did you find most interesting? \\
      Q4 & Which textbook was most helpful in aiding \\
         & your memorization or in solving questions?\\
      Q5 & How did you perceive the allocated \\
         & time for reading? \\
      \hline
    \end{tabular}
    \label{table:post_survey}
  \end{minipage}
\end{table*}

\section{Data Collection}
In this section, we introduce details about the participants' demographic information, experiment setup, and the experiment procedure.

\subsection{Participants}
This study recruited 24 participants from diverse national backgrounds, including Eastern Europe, South Asia, East Asia, the Middle East, South America, and North America. The participants, aged between 21 and 31 years with an average age of 27, were either university students or professionals based in Germany.
The General Data Protection Regulation (GDPR) required informed consent from all participants before the experiment.
The participants were then assigned to six groups in a fair and unbiased manner, each tasked with solving problems under varying conditions.

\begin{figure*}[t!]
  \centering
  \includegraphics[width=0.8\textwidth]{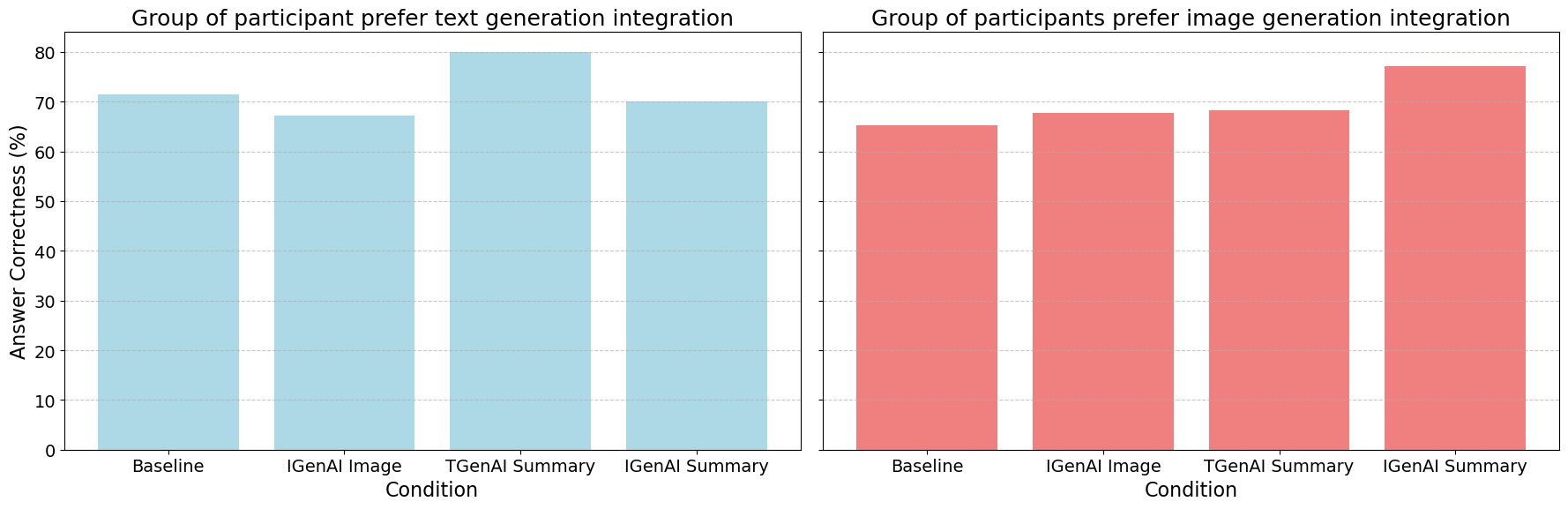}
  \caption{The comparison of post-reading test scores in solving questions between readers preferring text or image generation model.}
  \label{fig:llm_img}
\end{figure*}

\subsection{Experiment Setup}
During the story creation process, the following preference prompts were employed: by setting the configuration to nothing, animal, SF, and adventure, we ensured the diversity of the stories. 
Additionally, to avoid biases stemming from participants' prior knowledge, we generated entirely new stories that do not exist in the real world, thereby enabling fair comparisons. Each story was designed to be approximately 500 words in length.
In the question creation phase, the preference prompts were set to generate multiple-choice questions. The questions assessed comprehension, memorization, and synthesis, comprehensively evaluating the participants' reading tasks. The output was formatted as a JSON file to facilitate data processing.
Finally, for the summary creation phase, preference prompts generated summaries that were approximately 50 words in length. 

\subsection{Experiment Procedure}
Figure~\ref{fig:experiment_workflow} shows our experiment's overall experiment setup and workflow.
The Tobii Pro eye-tracker with a sampling rate of 90Hz, a remote device with an academic license that records eye movements, is mounted in the Microsoft Surface Studio 1.
All participants used identical desktop computers equipped with Microsoft Surface Studio 1.
Gen with $637.35 \times 438.90$ mm screen size, ensuring each story could be displayed on a single screen. 
The experiment was conducted in a quiet room.
This controlled environment ensured all participants experienced the experiment under the same conditions.
Participants first answered the consent form to confirm their participation in our data collection.
The consent form includes informing the subjects about data protection and compliance with the GDPR.

Once the participants confirmed their participation in this experiment, they answered the pre-survey.
Table~\ref{table:pre_survey} shows a list of questions asked in the pre-survey.
As explained in Section~\ref{subsubsec:group_selection}, we divided the participants into six groups to avoid bias in performance due to the difficulty of each story in this experiment.
Hence, we asked participants to indicate which group they would be involved in pre-survey Q1.
The participant answered the remaining questions as appropriate.
After the pre-survey, participants worked on Tobii Pro eye-tracker calibration.
Participants look at the dots on the screen to estimate the geometric characteristics of a participant's eyes.
Once calibration is done, we start eye-tracking data recordings.
Participants then access the webpage~\footnote{\url{https://generative-ai-textbooks.netlify.app/}}, our experiment web application.
The first page shows the selection of the group number, and the participants choose the one required by the experiment conductor.

Participants read four stories under different reading conditions (C1-C4).
Each group read the stories in a specific sequence to control for order effects.
The combination of story and reading conditions will vary from group to group.
Participants were informed before the reading time limit to standardize the time spent on each story.
The time limit for each story was adjusted according to the word count as explained in Section~\ref{subsubsec:readin_time_limit}.
Following the \citeauthor{atkinson1968human}, after reading each story, participants engaged in a distraction task~\cite{atkinson1968human} designed to clear their short-term memory and minimize any immediate recall effects.
This task involved solving fundamental arithmetic problems for one minute.
Once the distraction task is over, the participant answers ten post-reading tests related to the story the participant read.
Participants then repeat the reading task with different stories and conditions (C2-C4).
Once the participant completes all reading conditions (C1-C4), finish working with the web application.

Lastly, participants worked on the post-survey, whose questions are listed in Table~\ref{table:post_survey}.
We ask questions that can qualify the subjective feedback of which textbook (reading condition) supports participants' reading comprehension.

\begin{figure*}[t!]
  \centering
  \includegraphics[width=0.9\textwidth]{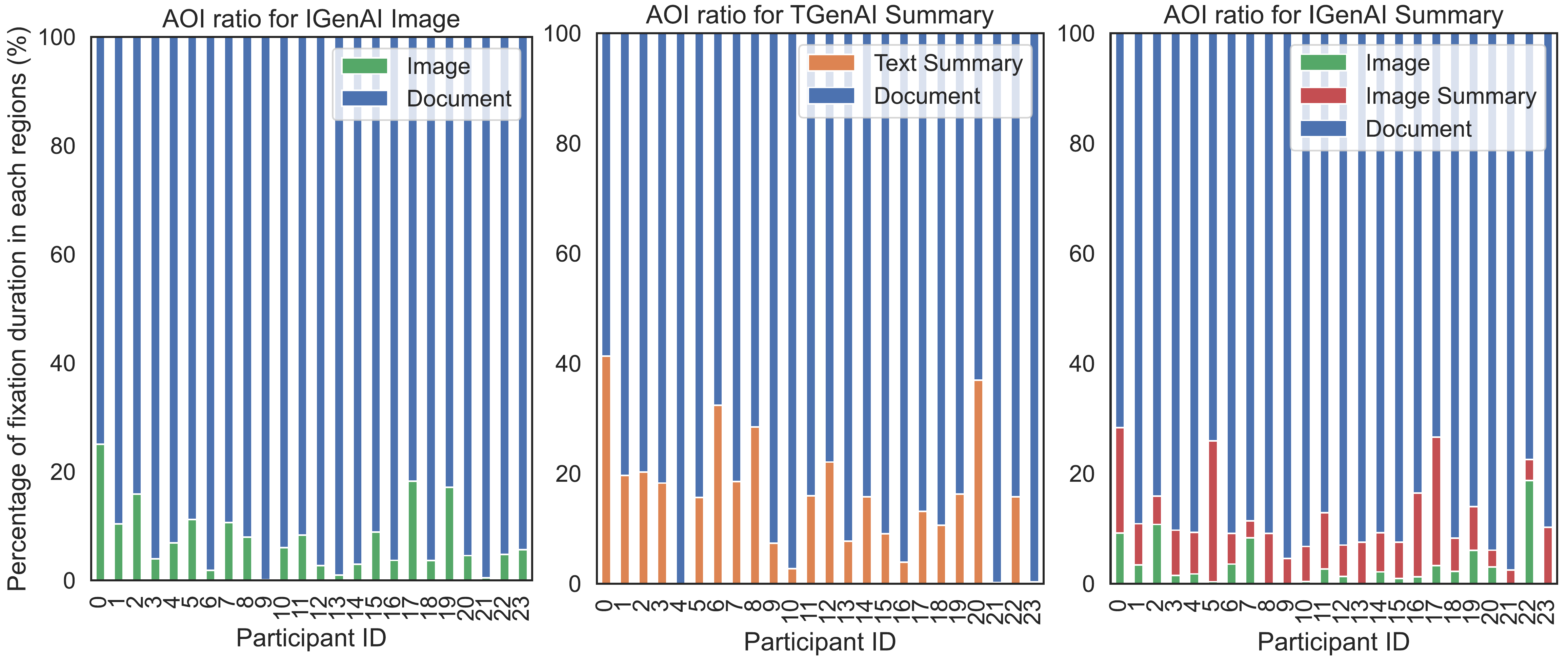}
  \caption{Comparison of the AOI ratio calculated by fixation duration for each participant in three different reading conditions (C2-C4). The ``Document'' represents the main reading text content. ``Image'', ``Text Summary'', and ``Image Summary'' represent the AOI ratio while looking at each area in the user interface prepared by IGenAI Image, TGenAI Summary, and IGenAI Summary.}
  \label{fig:percentage_fixation_duration_ratio}
\end{figure*}

\section{Result and Discussion}
\label{sec:result_and_discussion}
In this section, we present the results of our study and discuss their implications. We begin by examining overall reading comprehension across different conditions, followed by an analysis of the relationship between participants' preferences and their reading performance. We then explore eye movement patterns to understand how participants interacted with the GenAI materials.

\subsection{Overall Reading Comprehension Across Conditions}
\label{sec:res_overall}
Figure~\ref{fig:post_test_score_bargraph} shows the average post-reading test scores under different reading conditions. 
The \textit{IGenAI Image} condition (text with AI-generated images) yielded an average score 1.25\% higher than the \textit{Baseline} (text-only), suggesting a slight improvement in comprehension when images are included.
However, the \textit{IGenAI Image} condition exhibited greater variability in scores compared to the consistently high performance in the \textit{Baseline}, indicating that while some participants benefited from the images, others did not experience the same improvement.

Both GenAI summary conditions (\textit{TGenAI Summary} and \textit{IGenAI Summary}) significantly outperformed the \textit{Baseline}, highlighting that providing summaries—whether text-based or image-based—enhances reading comprehension. Notably, the \textit{IGenAI Summary} condition showed more variability in scores than the \textit{TGenAI Summary}, suggesting that image summaries are highly effective for some readers but less helpful for others.

These findings indicate that adding images as supplementary material introduces more variability in performance compared to adding text summaries. While images can enhance comprehension for some participants, they may also lead to inconsistent outcomes due to individual differences in processing visual information.

\subsection{Relationship Between Participants' Preferences and Reading Comprehension} \label{sec:res_int_ans}

Figure~\ref{fig:llm_img} illustrates the relationship between participants' preferences for learning material formats and their corresponding post-reading test scores.
Participants were grouped based on their preferred medium: text generation (TGenAI) and image generation (IGenAI).
This division allowed for a detailed analysis of how different types of supplementary materials impacted their comprehension and retention.

In the text-preference group (TGenAI), the highest average test scores were achieved under the \textit{TGenAI Summary} condition, demonstrating the benefit of providing a text-based overview. 
The \textit{Baseline} condition, which consisted of text-only content, yielded the second-highest scores.
In contrast, participants in this group performed noticeably worse under the \textit{IGenAI Summary} condition, and their lowest scores were recorded under the \textit{IGenAI Image} condition.
This suggests that visual summaries and images were less effective for participants favoring text-based materials.

Conversely, in the image-preference group (IGenAI), the \textit{IGenAI Summary} condition produced the highest test scores. 
Participants in this group also performed relatively well under the \textit{IGenAI Image} and \textit{TGenAI Summary} conditions. 
However, their lowest scores were observed in the \textit{Baseline} condition, indicating that the absence of visual aids hindered their comprehension and learning outcomes.

These findings highlight the importance of aligning supplementary educational materials with learners' preferences.
Participants favoring text generation achieved higher scores with TGenAI materials, particularly when supplemented with text summaries.
Similarly, those inclined toward image generation performed best with IGenAI materials, benefiting from both image summaries and standalone images.
These results suggest that tailoring educational content to individual preferences—textual or visual—not only improves engagement but also enhances comprehension, retention, and overall learning outcomes.

By leveraging personalized supplementary materials, this study underscores the potential of cognitive augmentation in education. Customizing content delivery based on learners' preferences can foster a more effective and enjoyable learning experience, paving the way for improved academic performance and deeper intellectual engagement.

\begin{figure*}[t!]
    \centering
    \begin{subfigure}[b]{0.46\textwidth}
        \centering
        \includegraphics[width=\textwidth]{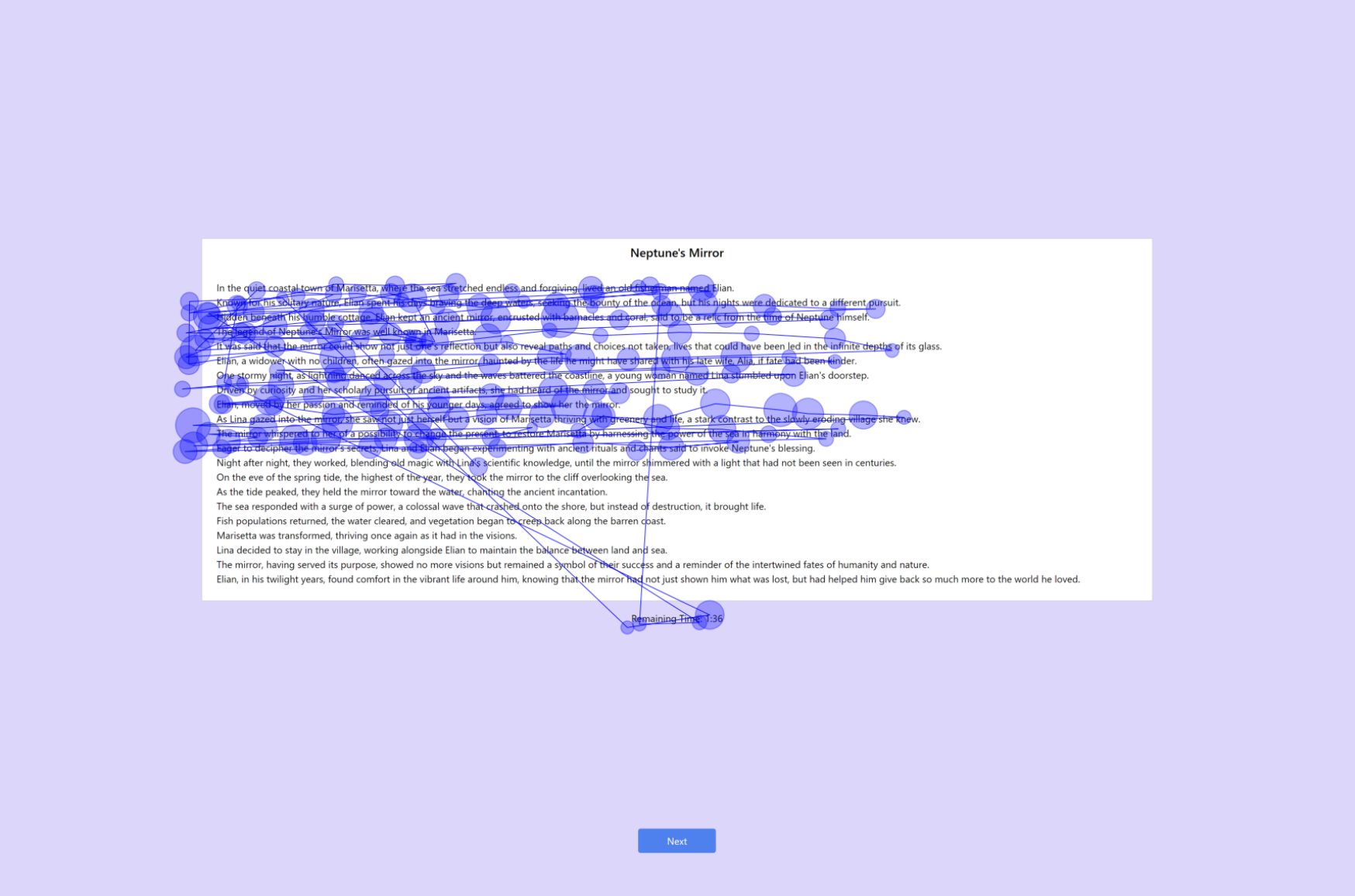}
        \caption{Reading baseline document (PID=11)}
        \label{fig:zentai1}
    \end{subfigure}
    \hspace{2mm}
    \begin{subfigure}[b]{0.453\textwidth}
        \centering
        \includegraphics[width=\textwidth]{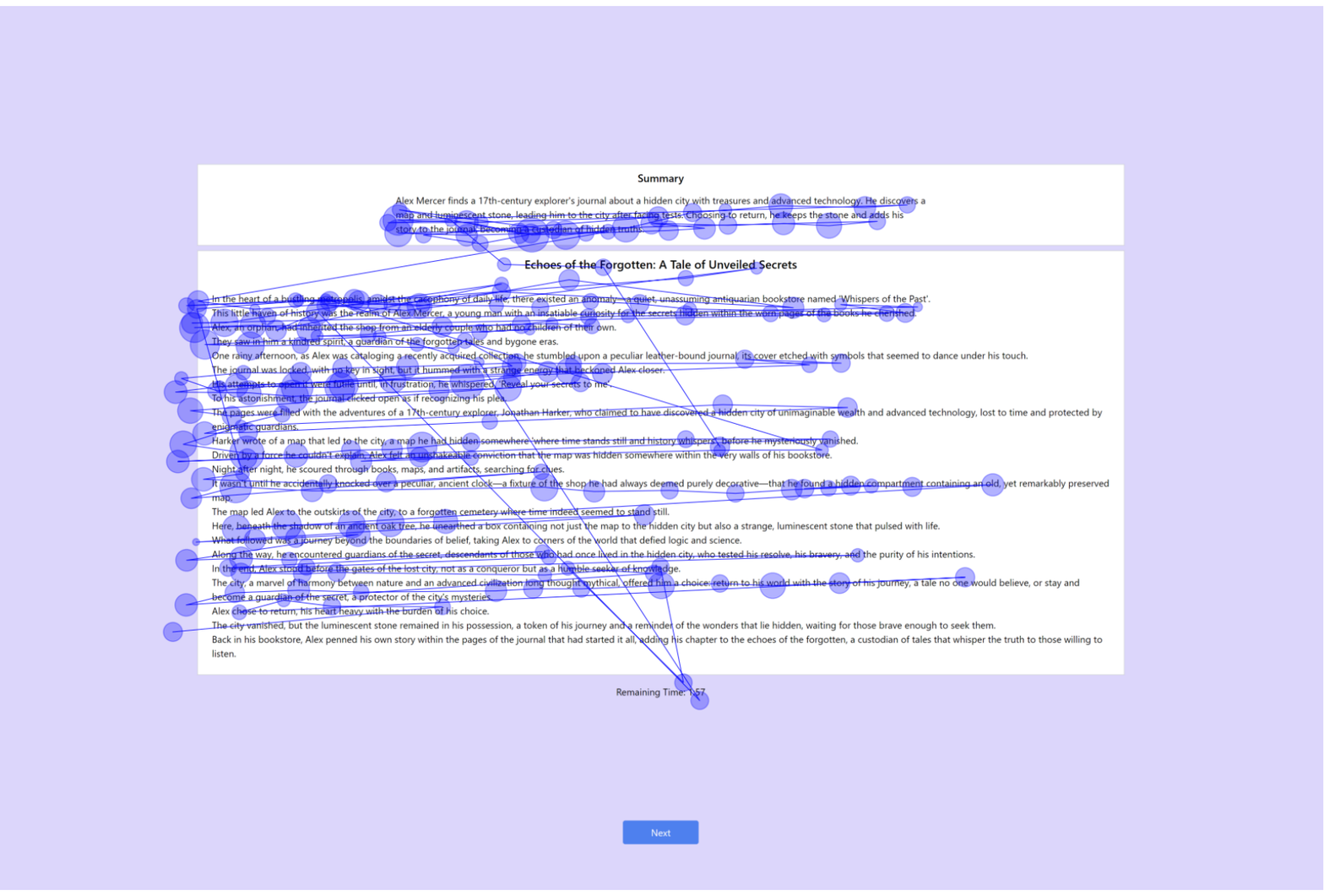}
        \caption{Reading TGenAI Summary (PID=11)}
        \label{fig:zentai2}
    \end{subfigure}
    \begin{subfigure}[b]{0.455\textwidth}
        \centering
        \includegraphics[width=\textwidth]{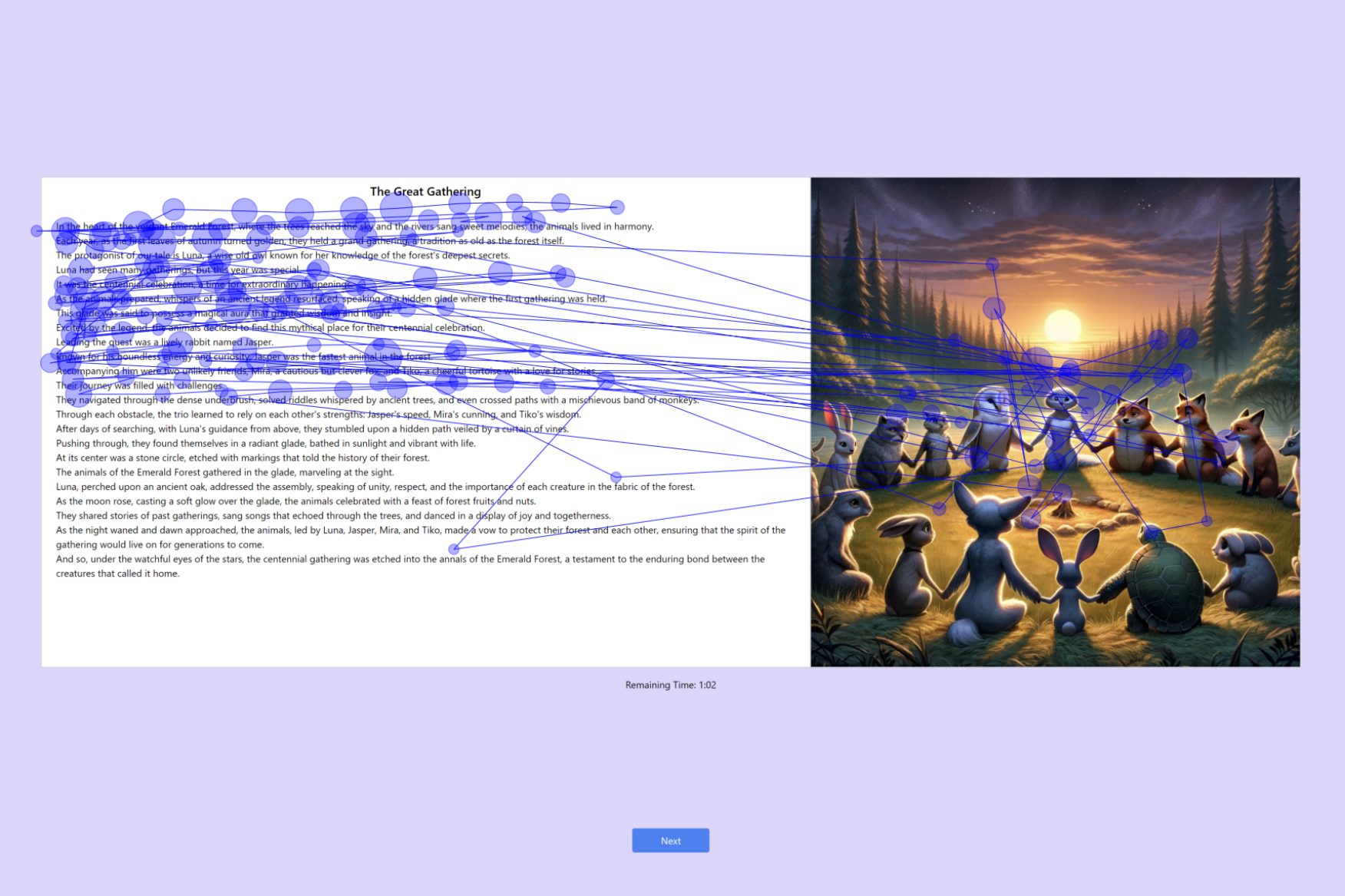}
        \caption{Reading IGenAI Image (PID=11)}
        \label{fig:zentai3}
    \end{subfigure}
    \hspace{2mm}
    \begin{subfigure}[b]{0.455\textwidth}
        \centering
        \includegraphics[width=\textwidth]{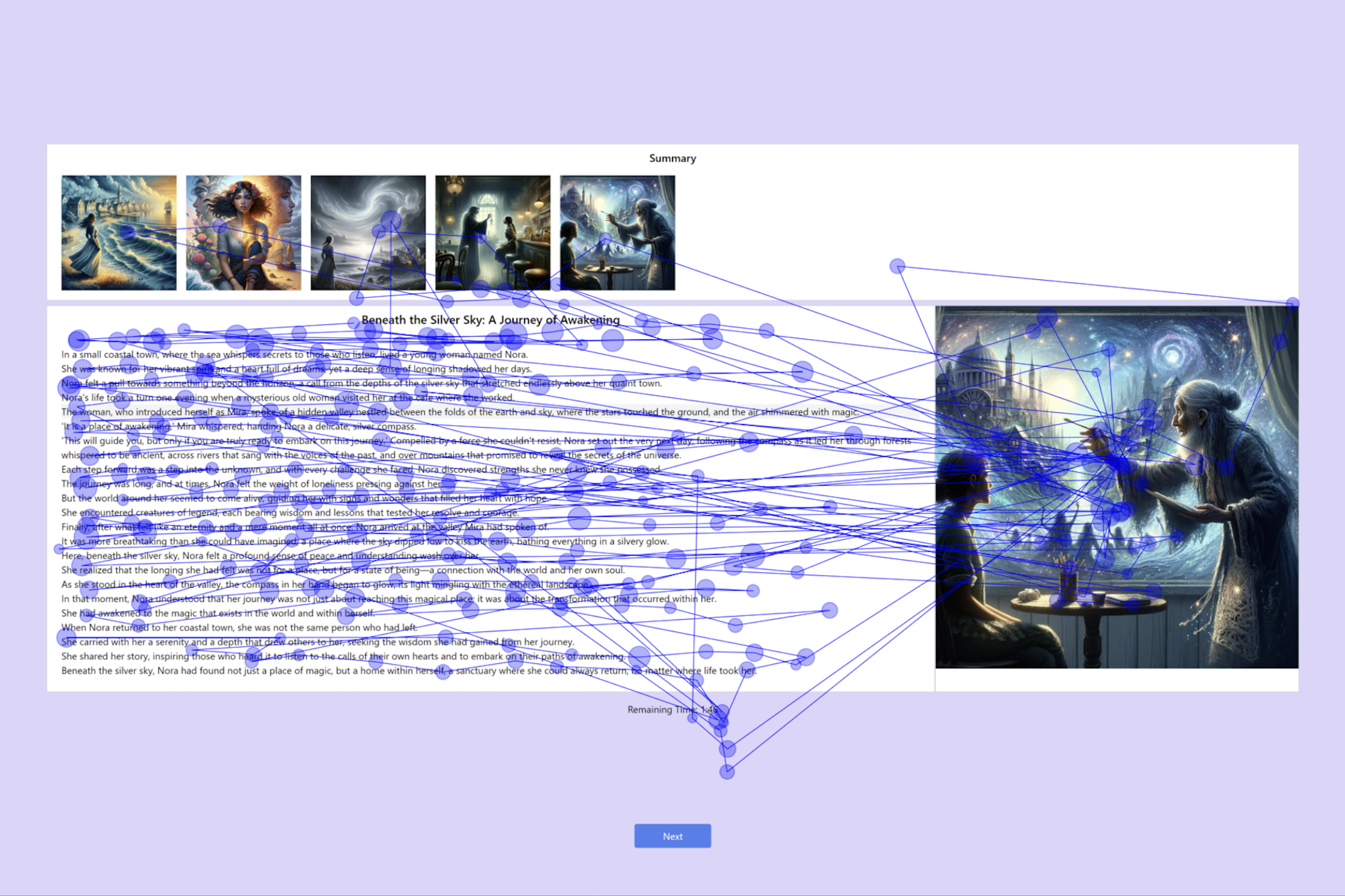}
        \caption{Reading IGenAI Summary (PID=11)}
        \label{fig:zentai4}
    \end{subfigure}
    \caption{Comparison of the gaze scan path for different reading conditions for ``Baseline'', ``TGenAI Summary'', ``IGenAI Image'' and ``IGenAI Summary''.}
    \label{fig:gen_ai_effect_of_summary}
\end{figure*}

\subsection{Eye Movement Patterns and Reading Behavior}
\label{sec:res_eye_ans}

We analyzed eye-tracking data to gain deeper insights into how participants engaged with the GenAI materials.

\subsubsection{Evaluation of AOI Ratio on Different Reading Conditions}
Figure~\ref{fig:percentage_fixation_duration_ratio} presents the AOI ratios, calculated by fixation duration, for each participant across three reading conditions (\textit{IGenAI Image}, \textit{TGenAI Summary}, and \textit{IGenAI Summary}). 
The results reveal distinct reading behaviors among participants. 
For example, \textit{Participant 17} demonstrated a strong preference for visual information, spending significant time viewing images in both IGenAI conditions.

In contrast, \textit{Participant 20} focused predominantly on the TGenAI Summary, indicating a preference for textual material. 
\textit{Participant 0} engaged extensively with all forms of supplementary information, while \textit{Participants 4} and \textit{23} showed disinterest in the Text Summary. 
\textit{Participant 9} largely ignored visual information, focusing on text, and \textit{Participant 21} exhibited minimal engagement with any GenAI material.
These observations illustrate that individuals process information differently, reflecting diverse cognitive strategies. 
GenAI summaries were generally more engaged than single images, possibly due to their comprehensive nature.
This underscores the importance of acknowledging individual differences in educational design and the potential benefits of personalized learning materials.

\subsubsection{Impact of Summaries on Reading Engagement}
Figure~\ref{fig:gen_ai_effect_of_summary} compares the gaze scan paths of a participant under different reading conditions. When supported by the TGenAI summary (Figures~\ref{fig:zentai1} and \ref{fig:zentai2}), the participant engaged with a wider range of document regions than in the baseline condition, suggesting that the textual summary facilitated more efficient navigation.

Including summaries significantly improved performance, as illustrated in Figure~\ref{fig:post_test_score_bargraph}. 
The presence of a summary allowed readers to grasp the story's overall structure early on, enabling quicker comprehension of subsequent content. 
Further analysis shows that both text and image summaries enhanced performance, though their impact varied based on reading speed. 
Text summaries required additional reading time, while image summaries provided an almost instant understanding of the story's flow, freeing up time for the main text and enhancing comprehension.

These results suggest that summaries function as an effective pre-reading tool for cognitive augmentation. 
In time-sensitive scenarios like this study, visual summaries outperformed text summaries due to their rapid processing advantage. 
The \textit{Summary Image Selector} developed in this research holds promise for improving AI-generated textbooks, particularly for speed-reading tasks, thereby fostering more efficient learning experiences.

\begin{figure*}[t!]
  \centering
  \begin{subfigure}[b]{0.48\textwidth}
      \centering
      \includegraphics[width=\textwidth]{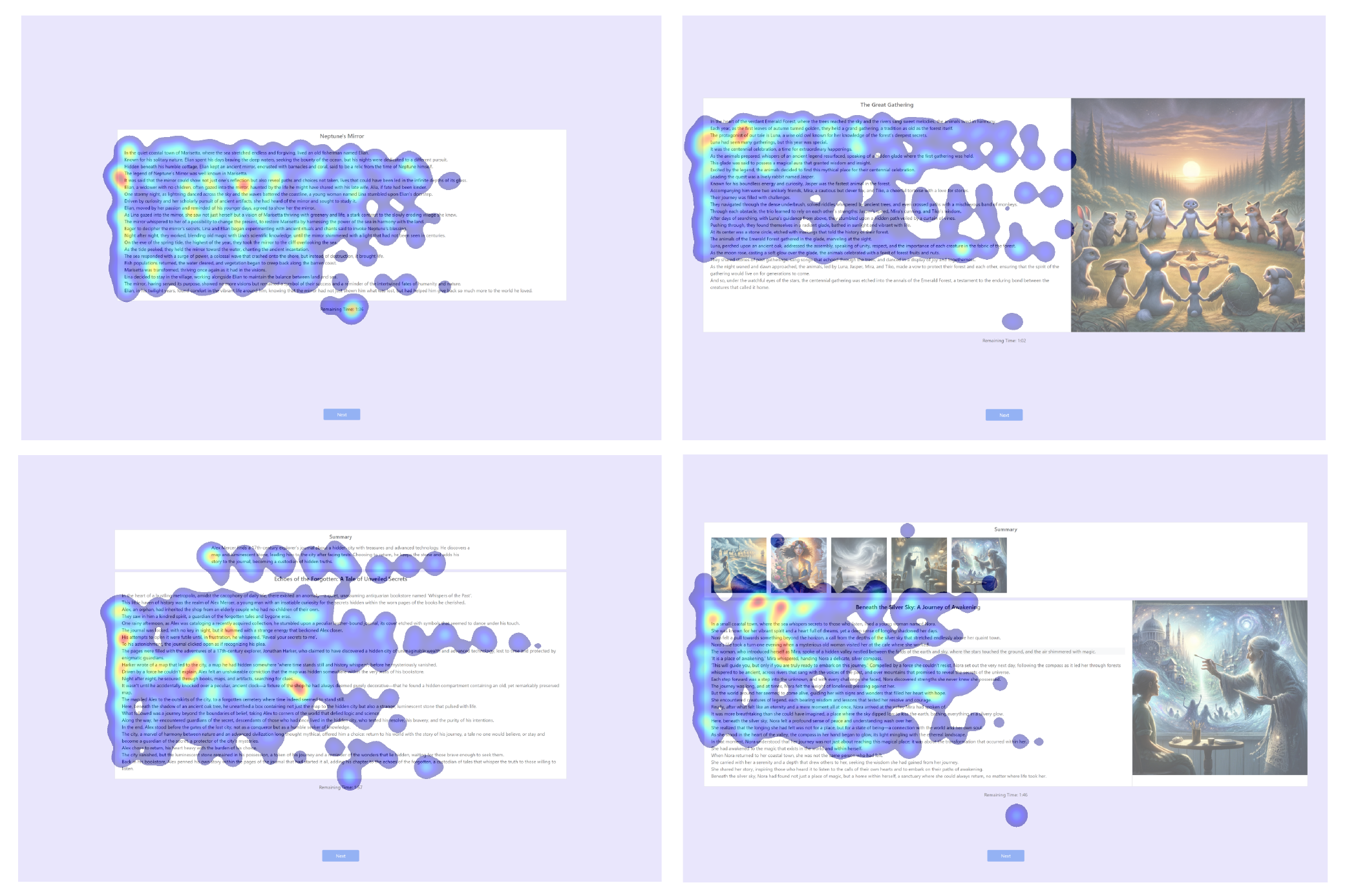}
      \caption{Gaze fixation heatmap of the verbal preference learner (PID=9)}
      \label{fig:heat_verbal}
  \end{subfigure}
  \hspace{2mm}
  \begin{subfigure}[b]{0.48\textwidth}
      \centering
      \includegraphics[width=\textwidth]{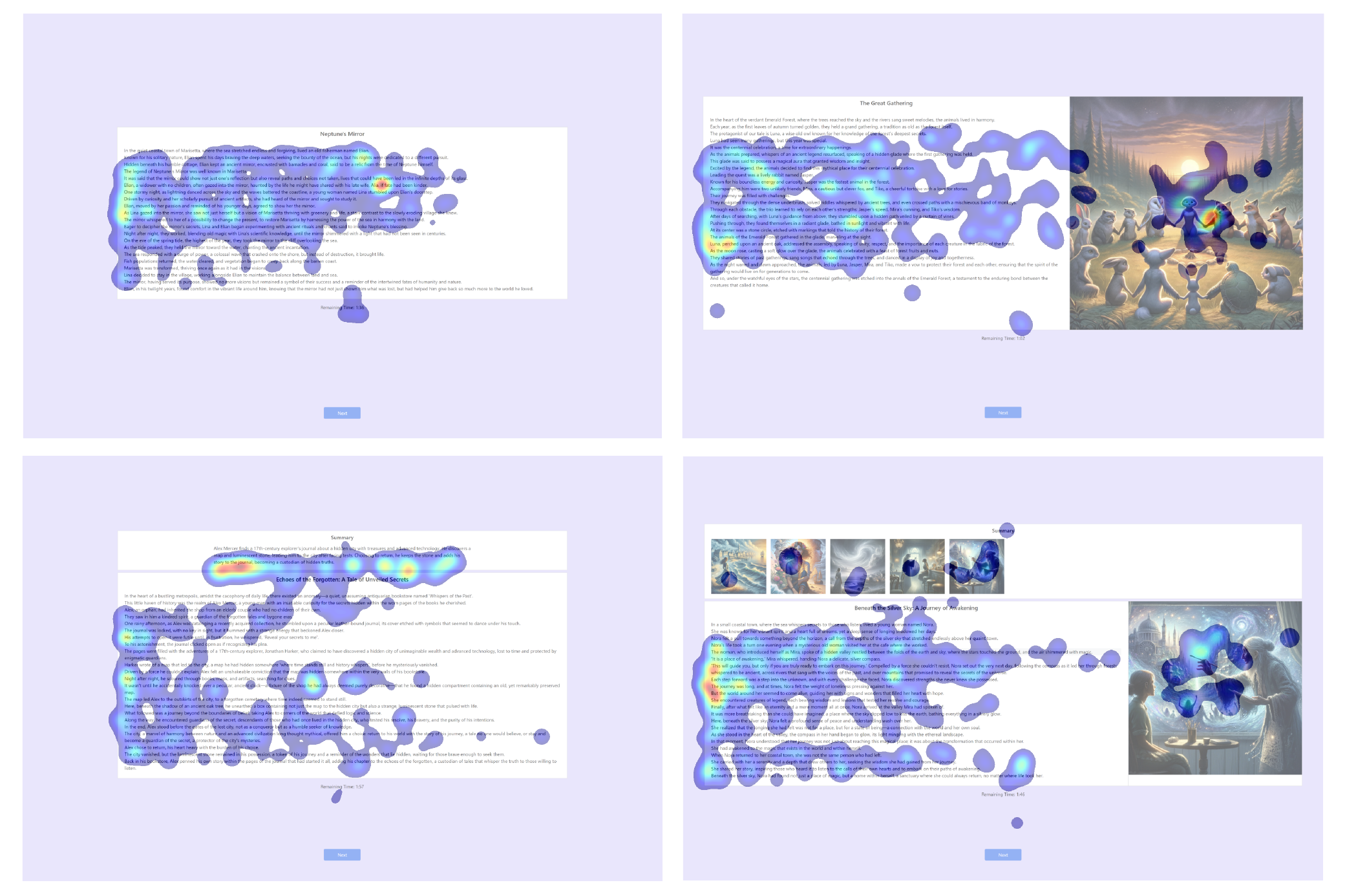}
      \caption{Gaze fixation heatmap of the visual preference learner (PID=0)}
      \label{fig:head_visual}
  \end{subfigure}
  \caption{The comparison with the reading behavior between verbal and visual preference learner. In the post-survey (Q3 and Q4), verbal preference learner is the sample of participants who chose ``text-generation'' as a preference for the reading condition, and visual preference learners chose ``image-generation'' as a preference for the reading condition.}
  \label{fig:heat_map}
\end{figure*}

\subsubsection{Verbal vs. Visual Preference Learners}
Based on participants' post-reading test scores and self-reported preferences, we observed a tendency suggesting that individual learning preferences might influence reading comprehension.
Some participants, termed \textit{Verbal Preference Learners}, appeared to process information more effectively through language and text, while others, \textit{Visual Preference Learners}, excelled when information was presented visually.

To explore the implications of these learning styles, we conducted further analysis.
While we cannot definitively claim a statistical correlation, the observed patterns indicate that personalization could enhance learning outcomes. 
We analyzed gaze heatmaps to illustrate these tendencies.
As shown in Figure~\ref{fig:heat_map}, \textit{Participant 19}, who preferred text generation, progressed through the text with minimal engagement with images, even when visual elements were present.
In contrast, participants who preferred image generation frequently alternated their gaze between text and images, indicating active integration of visual information.

These findings highlight the importance of incorporating both verbal and visual elements into educational materials to accommodate diverse learner preferences.
Our model demonstrates that personalization is possible by providing supplementary materials that align with individual cognitive styles.
By tailoring educational content to match learners' preferences, we can potentially enhance comprehension and retention across varied student populations.

\section{Limitations and Future Directions}
Our study indicates that AI-generated supplementary materials can significantly enhance post-reading test scores. However, several limitations remain, offering multiple avenues for further investigation.

\noindent
\textbf{Computational Bottleneck in Image Generation.}
The computational intensity of current diffusion-based models (e.g., DALL-E) poses scalability challenges, especially for real-time or large-scale educational contexts. Even smaller class settings may struggle with on-demand generation when bandwidth or hardware resources are limited. Future work might explore more efficient pipelines, hardware optimizations (e.g., GPU clusters), or specialized smaller models that maintain quality while reducing latency.

\noindent
\textbf{Domain-Specific Complexity.}
We focused on relatively simple narratives, for which the visual representations are straightforward. More specialized fields—such as chemistry, physics, or mathematics—require domain-specific visuals (e.g., chemical structures, equations) that exceed the capabilities of general-purpose generative models. Techniques like fine-tuning, advanced prompting, or hybrid approaches that combine symbolic knowledge with generative AI could improve the precision and realism needed in these complex subjects.

\noindent
\textbf{Cognitive Load and User Interface.}
Although supplementary images and summaries aided comprehension, additional on-screen elements led to more frequent gaze shifts (Figures~\ref{fig:zentai3} and \ref{fig:zentai4}), which some participants described as distracting. Future research might evaluate adaptive UI layouts—such as overlays or collapsible panels—that appear only when needed. Such adaptive interfaces could balance the benefits of multimodal information against the risk of overloading the learner’s visual attention.

\noindent
\textbf{Potential Bias in Participant Demographics and AI Familiarity.}
Our participants, largely young adults with prior experience using Large Language Models (LLMs) and Image Generation Models (IGMs), may not represent broader or less tech-savvy populations (see Figure~\ref{fig:llm_igm} in the supplementary material). Individuals with different cultural backgrounds, age ranges, or language proficiencies may respond differently to AI-augmented materials. Recruiting more diverse user groups—including older adults, K–12 students, or non-native English speakers with minimal AI exposure—can deepen understanding of how demographic factors modulate the effectiveness of generative textbooks.
\noindent
\textbf{Reliance on AI-Generated Stories and Questions.}
We employed LLM-generated narratives and comprehension questions to avoid relying on participants’ prior domain knowledge. However, these generated materials do not necessarily reflect the rigor or pedagogical structure of professionally authored textbooks or standardized tests. AI-generated questions may overemphasize factual recall at the expense of deeper inference. Future research can integrate established texts and validated question banks to better assess how generative AI translates to real-world educational settings.

\noindent
\textbf{Shallow Eye-Tracking Measures.}
Our study mainly considered fixation durations and areas of interest (AOIs), which capture broad viewing patterns. Additional psychophysiological or behavioral data, such as pupil dilation (an indicator of cognitive load), EEG signals (for workload), or changes in reading speed over time, might offer more nuanced insights into how AI-generated content shapes comprehension and engagement. Further work combining these measures with eye-tracking could refine our understanding of the mechanisms behind AI-enhanced reading.

\section{Conclusion}
This study demonstrates that incorporating AI-generated supplementary materials—such as images, text summaries, and image summaries—significantly enhances cognitive augmentation. 
Our findings indicate that learners' preferences affect reading behavior and post-reading scores, challenging the traditional one-size-fits-all approach to educational materials. 
Enhancing traditional text-based materials with AI-generated content can improve memory retention and comprehension. 
The study recruited 24 participants, and verified that integrating AI-generated supplementary materials significantly improved learning outcomes, increasing post-reading test scores by 7.50\%. 
As AI technology continues to evolve, integrating it into educational tools represents a powerful strategy for fostering learner engagement and understanding. 
This research highlights the potential of generative AI to create personalized educational experiences that enhance human intellect and cognitive augmentation.
Our study bridges gaps in traditional education and encourages the broader adoption of AI in personalized learning environments.

\begin{acks}
This work is partially supported by JSPS KAKENHI Grant Number 24K02962.
\end{acks}

\bibliographystyle{ACM-Reference-Format}
\bibliography{sample-base}

\end{document}